\IfFileExists{revtex4-2.cls}{
\newcommand*{\revtexcls}{revtex4-2}
}{
\newcommand*{\revtexcls}{revtex4-1}
}
\newcommand*{\ARXIV}{}
\ifdefined\ARXIV
\documentclass[twocolumn,amsmath,amssymb,showpacs,aps,longbibliography,superscriptaddress]{\revtexcls}
\else
\documentclass[twocolumn,amsmath,amssymb,showpacs,aps,longbibliography,superscriptaddress,prl]{\revtexcls}
\fi
\pdfoutput=1
\usepackage[bookmarks=false,linkcolor=blue,urlcolor=blue,colorlinks,citecolor=blue]{hyperref}
\usepackage{graphicx}
\usepackage{bm}
\usepackage{amssymb}
\usepackage{amsmath}
\usepackage{amsfonts}
\usepackage{mathtools}
\usepackage{nicefrac}
\usepackage{paralist}
\usepackage[rgb]{xcolor}
\usepackage{todonotes}
\usepackage{mathrsfs}
\usepackage{tabularx}
\usepackage{lipsum}
\usepackage{xspace}
\ifdefined\ARXIV
\else
    \usepackage{xr}
\fi
\setlipsumdefault{1-1}
\newcommand{\eg}{\emph{e.g.}\xspace}
\newcommand{\ie}{\emph{i.e.}\xspace}

\newcommand{\etc}{\emph{etc.}\xspace}

\newcommand{\wrt}{w.r.t.\xspace}

\newcommand{\iid}{\emph{i.i.d.}\xspace}
\def\multiset#1#2{\ensuremath{\left(\kern-.3em\left(\genfrac{}{}{0pt}{}{#1}{#2}\right)\kern-.3em\right)}}

\newtheorem{claim}{Claim}

\newcommand{\vv}{{\mathbf v}}

\newcommand{\s}{{\mathbf s}}
\newcommand{\e}{{\mathbf e}}

\newcommand{\HH}{{\mathcal H}}

\newcommand{\OO}{{\mathcal O}}

\newcommand{\C}{{\mathbb C}}
\DeclareMathOperator*{\EE}{{\mathbb E}}

\newcommand{\W}{{\mathcal W}}

\newcommand{\abs}[1]{\left\lvert#1 \right\rvert}

\DeclareMathOperator*{\argmin}{argmin}

\DeclareMathOperator{\RE}{Re}

\begin{document}
\title{Deep autoregressive models for the efficient variational \\ simulation of many-body quantum systems}

\author{Or Sharir}
\email{or.sharir@cs.huji.ac.il}
\affiliation{The Hebrew University of Jerusalem, Jerusalem, 9190401, Israel}
\author{Yoav Levine}
\email{yoavlevine@cs.huji.ac.il}
\affiliation{The Hebrew University of Jerusalem, Jerusalem, 9190401, Israel}
\author{Noam Wies}
\email{noam.wies@cs.huji.ac.il}
\affiliation{The Hebrew University of Jerusalem, Jerusalem, 9190401, Israel}
\author{Giuseppe Carleo}
\email{gcarleo@flatironinstitute.org}
\affiliation{Center for Computational Quantum Physics, Flatiron Institute, 162 5th Avenue, New York, NY 10010, USA}
\author{Amnon Shashua}
\email{shashua@cs.huji.ac.il}
\affiliation{The Hebrew University of Jerusalem, Jerusalem, 9190401, Israel}

\begin{abstract}
    Artificial Neural Networks were recently shown to be an efficient representation 
    of highly-entangled many-body quantum states. In practical applications, neural-network states
    inherit numerical schemes used in Variational Monte Carlo, most notably the use of
    Markov-Chain Monte-Carlo (MCMC) sampling to estimate quantum expectations. 
    The local stochastic sampling in MCMC caps the potential advantages of neural networks in two ways: (i) Its
    intrinsic computational cost sets stringent practical limits on the width and depth of the networks,
    and therefore limits their expressive capacity; (ii) Its difficulty in
    generating precise and uncorrelated samples can result in estimations of
    observables that are very far from their true value. Inspired by the state-of-the-art generative models used in machine learning, we propose a
    specialized Neural Network architecture that supports efficient and exact
    sampling, completely circumventing the need for Markov Chain sampling.
    We demonstrate our approach for two-dimensional interacting spin models, 
    showcasing the ability to obtain accurate results on larger system sizes than 
    those currently accessible to neural-network quantum states. 
\end{abstract}
\maketitle

\emph{Introduction.--} 
The theoretical understanding and modeling of interacting many-body quantum matter represents 
an outstanding challenge since the early days of quantum mechanics. 
At the heart of several problems in condensed matter, chemistry, nuclear matter, and more lies
the intrinsic difficulty of fully representing the many-body wave-function,
in principle needed to exactly solve Schrodinger's equation. 
These mainly fall into two categories: on one hand, there are states traditionally used in stochastic Variational Monte Carlo (VMC) calculations~\cite{mcmillan_ground_1965}. Chief example are Jastrow wave-functions \cite{jastrow_many-body_1955}, carrying high entanglement, but also with a limited variational freedom.
On the other hand, more recently, tensor-network approaches have been put forward, 
based on non-stochastic variational optimization, and most chiefly on entanglement-limited variational wave-functions~\cite{fannes1992finitely,perez2007matrix,
verstraete2004renormalization,vidal2008class}.

In an attempt to circumvent the limitations of the approaches above, 
architectures based on Artificial Neural Networks~(ANN) were proposed as variational wave
functions~\cite{Carleo:2017cn}. Restricted Boltzmann machines (RBM), which
represent relatively veteran machine learning constructs, were shown to be
capable of representing volume-law entanglement scaling in
2D~\cite{deng2017quantum,PhysRevB.97.085104,PhysRevX.8.011006,kaubruegger_chiral_2018}. 
Recently, other neural-network architectures have been explored. 
Most notably, convolutional neural networks (ConvNets) -- leading
deep learning architectures that stand at the forefront of empirical successes
in various Artificial Intelligence domains -- have been applied to both bosonic~\cite{saito_machine_2017} 
and frustrated spin systems ~\cite{choo_symmetries_2018}.

Despite the provable theoretical advantage of ConvNet architectures~\cite{levine2019quantum}, however, early numerical studies have been limited to relatively shallow architectures, far from the very deep networks used in modern machine learning applications.
This practical limitation is mostly due to two main factors. First, it is computationally expensive to obtain 
quantum expectation values over ConvNet states using stochastic sampling based on Markov Chain Monte Carlo (MCMC), as is it is customary in VMC applications. 
Second, there is an intrinsic optimization bottleneck to be faced when dealing with a large number 
of parameters. However, both limitations are routinely faced when learning deep autoregressive-models, recently introduced machine-learning techniques that have enabled previously intractable applications. 

In this paper, we propose a pivotal shift in the use of Neural-network Quantum States (NQS) for many-body quantum systems, 
that markedly sets a discontinuity with traditionally adopted VMC methods. Inspired by the latest advances 
in generative machine learning models, we introduce variational states for which both the sampling
and the optimization issues are substantially alleviated. Our model is composed of a ConvNet that allows direct, efficient, and \iid sampling from the highly entangled wave function it represents. 
The network architecture draws upon successful autoregressive models for representing and 
sampling from probability distributions. Those are widely employed in the machine learning
literature~\citep{NADE}, and have been recently used for statistical mechanics applications~\cite{wu_solving_2018},
as well as density matrix reconstructions from experimental quantum systems~\citep{Carrasquilla:2019wm}.
We generalize these autoregressive models to treat complex-valued wave-functions, obtaining highly
expressive architectures parametrizing an automatically normalized many-body
quantum wave-function.

\emph{Neural Autoregressive Quantum States.--} We consider in the following
a pure quantum system,
constituted by $N$ discrete degrees of freedom ${\s{\equiv}(s_1,\ldots,s_N)}$ (\eg spins, occupation numbers, \etc) 
such that the wave-function amplitudes $\Psi(s)$ fully specify its state. 
Here we follow the approach introduced in \cite{Carleo:2017cn}, 
and represent $\ln(\Psi(\s))$ as a feed-forward ANN, parametrized by a possibly large number of network connections. 
Given an arbitrary set of quantum numbers, $s$, the output value computation of the corresponding NQS, known as its \emph{forward pass}, can generally be described as a sequence of $K$ matrix-vector multiplications separated by the applications
of a non-linear element-wise \emph{activation} function $\sigma{:}\C{\to}\C$. More formally, 
the unnormalized log amplitudes are given by 
\begin{align}\label{eq:nqs}
    \ln(\Psi(\s)) &= W_K \sigma\left(W_{k-1} \sigma\left( \cdots \sigma\left(W_1 \s \right) \right) \right),
\end{align}
where $\W{\equiv}\left\{W_i{\in}\C^{r_i{\times}r_{i-1}}\right\}_{i=1}^K$,
$r_0{=}N,r_K{=}1,r_1,\ldots,r_{K-1}$ are known as the \emph{widths}
of the network, and $K$ as the \emph{depth}. In practice, specialized
variants of eq.~\ref{eq:nqs} are commonly used, \eg early applications have 
focused on shallow architectures ($k{=}1$) such as Restricted Boltzmann Machines, for which 
the activation function is typically taken to be $\sigma(z){=}\ln\cosh(z)$. 
Other, deeper, choices are often advantageous, such as convolutional networks, in which most 
of the matrices are restricted to act on a subset of the quantum numbers, computing convolutions with small filters.

Given a NQS representation of a many-body quantum state, estimating physical observables $\langle \Psi |\OO|\Psi\rangle$ of a local operator $\OO$, is in general analytically intractable, but can be realized numerically through a stochastic procedure,
as done in VMC. 
Specifically, $\langle \Psi |\OO|\Psi\rangle{=}\langle O^{\mathrm{loc}}\rangle_{\mathcal{P}}$, where
$\langle {\dots} \rangle_{\mathcal{P}}$ denote statistical expectation values over
the Born probability density $\mathcal{P}(\s){\equiv}\abs{\Psi(\s)}^2$, and 
$O^{\mathrm{loc}}{\equiv}\sum_{\s^\prime} \langle \s | \mathcal{O}| \s^\prime\rangle \Psi(\s^\prime)/\Psi(\s)$ is the corresponding
statistical estimator. 
In the vast majority of VMC applications, including NQS so-far, a MCMC
algorithm is typically used to generate samples from $\mathcal{P}(\s)$. While MCMC is a rather 
flexible technique, it comes with a large
computational cost, especially for deep ANNs. 
Additionally, though MCMC asymptotically generates samples that are correctly distributed, in
practice it can be plagued by very large autocorrelation times, and lack of ergodicity, that can 
severely affect the quality of the samples being generated. 

In light of these limitations, we propose here a specialized network
architecture that instead supports efficient and exact sampling. 
Our approach is an extension of Neural Autoregressive Density
Estimators~(NADE)~\citep{NADE} to quantum applications, resulting in what we dub \emph{Neural
Autoregressive Quantum States}~(NAQS). To start with, first consider the task of 
representing a probability distribution with NADE models. These models build on the so-called
 autoregressive property, which entails a  
decomposition of the full probability distribution as a product of conditionals, \ie 
$P(s_1,\ldots,s_N){=}\prod_{i=1}^N p_i(s_i | s_{i-1},\ldots, s_1)$. 
The power of these models comes from the observation that, for every $i$, 
the conditional probabilities $p_i$ can be
individually represented as an ANN receiving as input the variables $s_1,\ldots,s_{i-1}$
and outputting a vector $\vv_i{\equiv}(v_{i,s_1},v_{i,s_2},{\dots},v_{i,s_M})$ representing the unnormalized probabilities for~$s_i$ to take one of the $M$ possible discrete values $s_{j}$, conditioned on given $s_1,\ldots,s_{i-1}$. It is
crucial that each output vector $\vv_i$ does not depend on the value of
$s_i$ or any of the variables appearing with a larger index, $s_{i+1},\ldots,s_N$, for
a pre-chosen ordering. To ensure that each network outputs a valid
conditional distribution, it is then sufficient to take the exponent of each
entry and normalizing it according to the $l_1$ norm, \ie
$p_i(s_i | s_{i-1},\ldots, s_1){=}{\exp(v_{i,s_i})}{/}{\sum_{s^\prime} \abs{\exp(v_{i,s^\prime})}}$, also known as a \emph{Softmax} operation.

Even though it is possible to use $N$ separate networks for each of the $N$
conditional probabilities, and each accepting a variable number 
of inputs, in practice it is more common to use a single ANN
that accepts $N$ inputs and outputs $N$ probability vectors. In this case, the
autoregressive property is enforced by masking the
inputs $s_i,\ldots,s_N$ for the $i$'th output vector, \ie ensuring that the contributions
of higher-ordered spins to the output of the network vanish.
PixelCNN~\citep{van2016conditional} is such an architecture, and is built as a sequence of
\emph{masked} convolutional layers, whose filters are restricted to having zeros
at positions ``ahead''. For example, in a one dimensional system, a filter of width $R$, 
where $R$ is odd, would be constrained to have
$(w_1, \ldots, w_{\nicefrac{(R-1)}{2}}, 0, \ldots, 0)$, and thus the $i$th
output of each layer depends uniquely on the indices at $s_1,\ldots,s_{i-1}$.

A chief advantage of networks with the autoregressive property, is that directly drawing 
samples according to $P(\s)$ is conceptually straightforward. 
One can sample each $s_i$ in sequence, according to its given conditional probability
that depends just on the previously sampled $(s_1,\ldots,s_{i-1})$. 
Carefully exploiting the intrinsic sparseness of the network weights, further leads to 
a very efficient algorithm for sampling~\citep{ramachandran2017fast}. Remarkably, the complexity of sampling a
full string $s_1\dots s_N$ in a PixelCNN architecture can be reduced to the complexity of just a single forward pass.

\begin{figure*}
    \centering
    \includegraphics[width=1\linewidth]{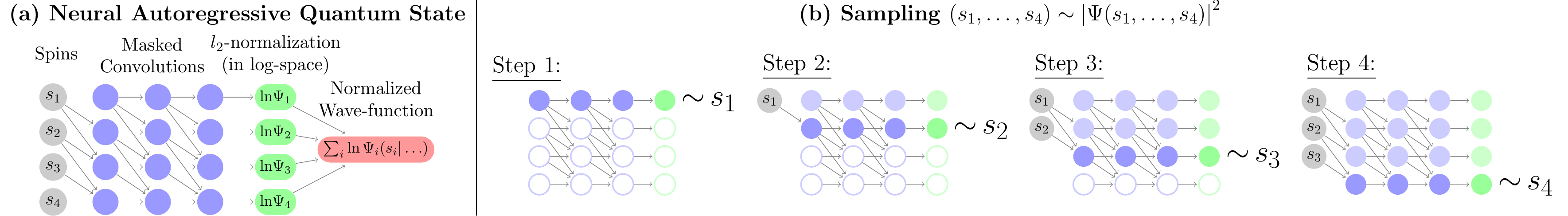}
    \caption{\label{fig:naqs}Neural Autoregressive Quantum States are neural networks that
    represent a normalized wave-function, $\Psi(s_1,\ldots,s_N)$, by factoring it to a sequence
    of normalized \emph{conditional} wave-functions, denoted by $\Psi_i(s_i|s_{i-1},\ldots,s_1)$
    for the $i$'th particle, in a manner similar to that of Neural Autoregressive Density Estimator
    (see eq.~\ref{eq:naqs_final}). 
    \textbf{(a)} Illustration of a deep 1D-convolutional NAQS model following the
    PixelCNN~\citep{van2016conditional} architecture. Each column of nodes represent a layer
    in the network, starting with the input layer representing the $N$-particle configuration
    $(s_1,\ldots,s_N)$. Each internal node in the graph is a complex vector computed according
    to its layer type. Namely, masked convolutions are limited to having local connectivity,
    where a node at the $j$'th row is only connected to nodes with
    connections to $s_i$ where $i{<}j$. All inputs to a node at the $l$'th layer are multiplied
    by a matrix $W^{(l)}$, shared across all rows in the same layer, and followed by applying
    a non-linear element-wise function $\sigma{:}\C{\to}\C$.
    \textbf{(b)} Depicts the exact sampling algorithm for NAQS, where empty nodes represent
    unused nodes, and filled but faded nodes represent cached results from previous steps.
    The quantum number of each particle is generated sequentially, by computing its respective conditional wave-function,
    and sampling according to the squared magnitude. Notice that only a single row is processed at
    each step, and so sampling a complete configuration has the same runtime as a single forward pass.}
\end{figure*}

Our NAQS model for representing wave-functions is
based on the same NADE principles so-far described.
Specifically, just as probability
functions can be factorized into a product of conditional probabilities, we
represent a normalized wave-function as a product of normalized
\emph{conditional} wave-functions, such that 
\begin{align}\label{eq:naqs}
\Psi(s_1,\ldots,s_N){=}\prod_{i=1}^N \psi_i(s_i | s_{i-1}, \ldots, s_1),
\end{align}
where $\psi_i(s_i | s_{i-1},\ldots,s_1)$ are such that, for any fixed
$(s_1,\ldots,s_{i-1}){\in}\left\{1,\ldots,M\right\}^{i-1}$, they satisfy the normalization condition
$\sum_{s^\prime} \abs{\psi_i(s^\prime | s_{i-1}, \ldots, s_1)}^2{=}1$. If this condition holds, then a strong normalization condition for the full wave-function follows~(see app.~\ref{app:proof} for proof):
\begin{claim}\label{claim:naqs_norm}
    Let $\Psi{:}[M]^N{\to}\C$ such that $\Psi(s_1,\ldots,s_N){=}\prod_{i=1}^N \psi_i(s_i | s_{i-1}, \ldots, s_1)$,
    where $\left\{\psi_i\right\}_{i=1}^N$ are normalized conditional wave-functions. Then,
    $\Psi$ is normalized, \ie, $\sum_{s_1,\ldots,s_N} \abs{\Psi(s_1,\ldots,s_N)}^2{=}1$. 
\end{claim}
As in the NADE case, we represent conditional wave-function with an ANN
accepting $(s_1, \ldots, s_{i-1})$ and outputting a complex vector
$\vv_i{\equiv}(v_{i,s_1},v_{i,s_2},\ldots,v_{i,s_M}){\in}\C^M$ for each of the $M$ 
possible values taken by the local quantum numbers $s_i$. To obtain a normalized
conditional wave-function, we take its exponent and normalize it according to the $l_2$-norm, \ie,
$\psi_{i}(s_i | s_{i-1}, \ldots, s_1)\,{=}\,\hat{v}_{i,s_i}\,{\equiv}\,{\exp(v_{i,s_i})}/{\sqrt{\sum_{s\prime} \abs{\exp(v_{i,s^\prime})}^2}}$.
Given this parametrization, the full wave-function log-amplitude $\ln\Psi(s_1\dots s_N)$ is easily obtained, once all the vectors $\vv_1,\ldots,\vv_N$ have been computed, as given by:
\begin{align}\label{eq:naqs_final}
    \ln \Psi(\s)&{=}\sum_{i=1}^N \left(v_{i,s_i} - \frac{1}{2}\ln \sum_{s^\prime} \abs{\exp(v_{i,s^\prime})}^2\right).
\end{align}
As in the probabilistic autoregressive model, we can represent the entire NAQS by a single neural
network outputting $N$ complex vectors, as illustrated in Fig.~\ref{fig:naqs}a. Though our
proposed architecture can work with either complex- or real-parameters, we have found that
using the latter work better, where we represent each complex conditional log-amplitude using two real values, log-magnitude and phase.

Moreover, there is a special relationship between a NAQS and its induced
Born probability, since $\abs{\Psi(s_1,\ldots,s_N)}^2{=}\prod_{i=1}^N \abs{\psi_i(s_i | s_{i-1}, \ldots, s_1)}^2$,
implying that $|\psi_i(s)|^2$ is a valid conditional probability. Thus, the induced Born probability of a NAQS has
the exact same structure of a NADE model. Specifically, taking the squared magnitude
of its output vectors, \ie, $\forall{i,s^\prime},\bar{v}_{i,s^\prime}{=}\abs{\hat{v}_{i,s^\prime}}^2$, transform
NAQS into a standard NADE representation of this distribution, which importantly includes
its efficient and exact sampling method. In contrast to standard MCMC sampling
employed for correlated wave-functions, NAQS thus allows for direct, efficient sampling with the
computational complexity of a 
single forward pass, as depicted in Fig.~\ref{fig:naqs}b.

\emph{Optimization.--} The NAQS representation of many-body wave functions can be
used in practice for several applications. These include for example ground-state search~\cite{Carleo:2017cn}, quantum-state tomography~\cite{torlai_neural-network_2018}, dynamics~\cite{Carleo:2017cn}, and quantum circuits simulation~\cite{jonsson_neural-network_2018}. 
Here we more specifically focus on the task of finding the ground state of a
given Hamiltonian $\HH$. In this context, we denote by $\Psi_\W$ the wave-function represented
by a NAQS of a fixed architecture that is parameterized by $\W$, and we wish
to find $\W$ values that minimize the energy, \ie, $\W^*{=}\argmin_\W E(\W)$, where
$E(\W){\equiv}\langle \Psi_\W \vert \HH \vert \Psi_\W \rangle{=}\EE_{\s \sim \abs{\Psi_\W}^2}[E_{\mathrm{loc}}(\s;\W)]$,
$E_{\mathrm{loc}}(\s;\W){\equiv}\sum_{\s'} \HH_{\s, \s'} \frac{\Psi_\W(\s')}{\Psi_\W(\s)}$,
and $\HH$ is usually a highly sparse matrix, and so computing $E_{\mathrm{loc}}$ for
a given sample takes at most $O(N)$ forward passes.

The common approach for solving the optimization problem above with an NQS is
to estimate the gradient of $E(\W)$ with respect to $\W$, and use variants of
stochastic gradient descent~(SGD) to find the minimizer of $E(\W)$. Estimating the gradient
can be done by first employing a variant of the log-derivative trick, \ie,
\begin{align}\label{eq:gradient_est}
    \frac{\partial E}{\partial \W} &{=} \EE_{\s \sim \abs{\Psi_\W}^2}\left[2\RE\left(\left(E_{\mathrm{loc}}(\s)^* - E^* \right) \frac{\partial \ln \Psi_\W}{\partial \W}\right)\right].
\end{align}

Now, while we can efficiently compute the log derivative of $\Psi_\W$,
exactly computing the expected value is intractable, but we can still approximate
it by computing its value over a finite batch of samples $\{\s_{(i)}\}_{i=1}^B$.
The quality of this approximation depends on the batch size, $B$,
but also on the degree of correlations between the individual samples. 
The advantages of our direct sampling method supported by NAQS over MCMC 
are twofold in this context: (i)~Faster sampling: each individual sample can be generated with fewer network passes,
and generating a batch of samples is embarrassingly parallel,
as opposed to the sequential nature of MCMC; (ii)~Faster convergence: because the generated samples are
exact and \iid, and so result in more accurate estimates of the gradient
at each step.

\emph{Experiments.--} As a first benchmark for our approach, 
we consider a case where MCMC sampling can be strongly biased. 
A paradigmatic quantum system exhibiting this issue is found in the ferromagnetic phase of
the transverse field Ising model. 
The Hamiltonian for this model is given by
$H{=}{-}J\sum_{<i,j>}\sigma_z^i\sigma_z^j{-}\Gamma \sum_{i}\sigma_x^i$,
where the summation runs over pairs of lattice edges. Here we study the case of a
2D square lattice with open boundary conditions, and for varying strengths of
the transverse field. The system is in a ferromagnetic phase when the transverse
magnetic field $\Gamma$ is weak with respect to the coupling constant, and
specifically in 2D when $\Gamma{<}\Gamma_c{\simeq}3.044J$ \cite{blote_cluster_2002}. 

\begin{table}
\begin{tabular}{ccccc}
$\Gamma$ & NAQS Energy & QMC Energy & NAQS $\left\langle\abs{\sigma_z}\right\rangle$ & QMC $\left\langle\abs{\sigma_z}\right\rangle$   \\ \hline
2.0J   & -2.4096022(2) & -2.40960(3) & 0.78326(2) & 0.78277(38) \\
2.5J & -2.7476550(5) & -2.74760(3) & 0.57572(3) & 0.57566(63) \\
3.0J   & -3.1739005(5) & -3.17388(4) & 0.16179(4) & 0.16207(54) \\
3.5J & -3.6424799(3) & -3.64243(4) & 0.11094(3) & 0.11011(30) \\
4.0J   & -4.1217979(2) & -4.12178(4) & 0.09725(2) & 0.09728(24)
\end{tabular}
\caption{\label{table:ground_state_values}
     Estimates of the ground state energies of the transverse-field Ising model for
     different values of $\Gamma$ on a $12{\times}12$ lattice, and the corresponding
     estimates of $\left\langle\sigma_z\right\rangle$, as obtained by either NAQS or QMC.
}
\end{table}

\begin{figure}
    \centering
    \includegraphics[width=\linewidth]{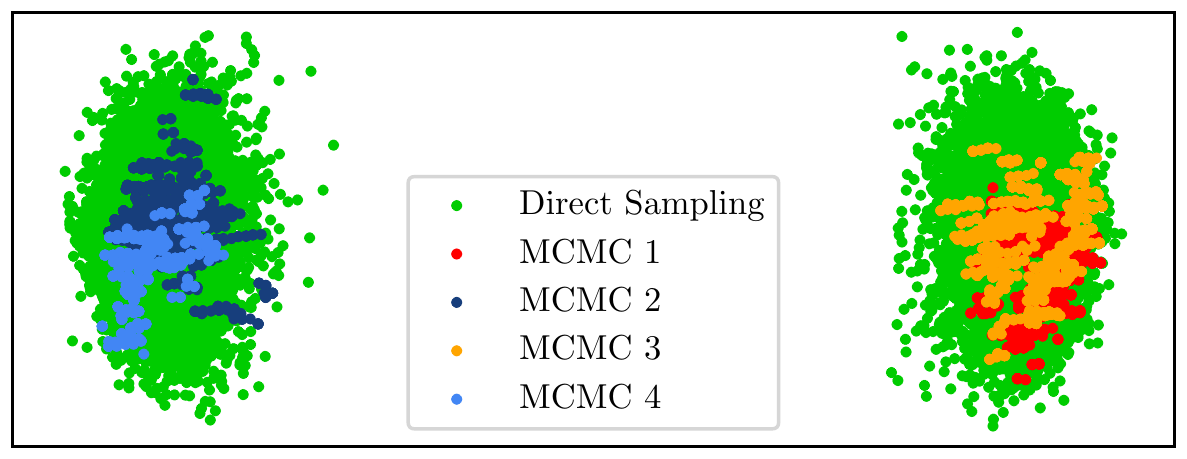}
    \caption{\label{fig:broken_sym}
    An illustration of the two modes of the ground state, by taking the first two
    principal components of the generated samples. The green points correspond to our direct
    sampling method, and the other colors represent different MCMC chains.
    The plot was generated by training a NAQS on the transverse-field Ising model with $\Gamma{=}2J$,
    below the critical value, on a $12{\times}12$ lattice until convergence
    to the ground state, and then sampling from the trained NAQS using either
    our direct sampling method, or 4 separate MCMC samplers.
    }
\end{figure}

In order to verify the correctness of the model proposed in section~2, we begin
by comparing the ground state energy and system magnetization obtained
for a $12{\times}12$ system with those obtained by an unbiased quantum Monte Carlo
(QMC) simulation. Using our open-source library, FlowKet~\footnote{FlowKet:
an open-source library based on Tensorflow for running Variational
Monte-Carlo simulations on GPUs,
\url{https://github.com/HUJI-Deep/FlowKet}}, we employ a NAQS model
following the PixelCNN architecture, using
the ADAM~\citep{kingma2014adam} SGD variant with the gradient estimator of
eq.~\ref{eq:gradient_est}. Additional technical details are listed in app.~\ref{app:tech}. 
Table~\ref{table:ground_state_values} shows that our model achieves
very high accuracy for both magnetization and energy densities for different transverse
field values across the phase diagram: when the system is in the ferromagnetic phase,
the normal phase, and near the phase transition.

\begin{figure}
    \centering
    \includegraphics[width=1\linewidth]{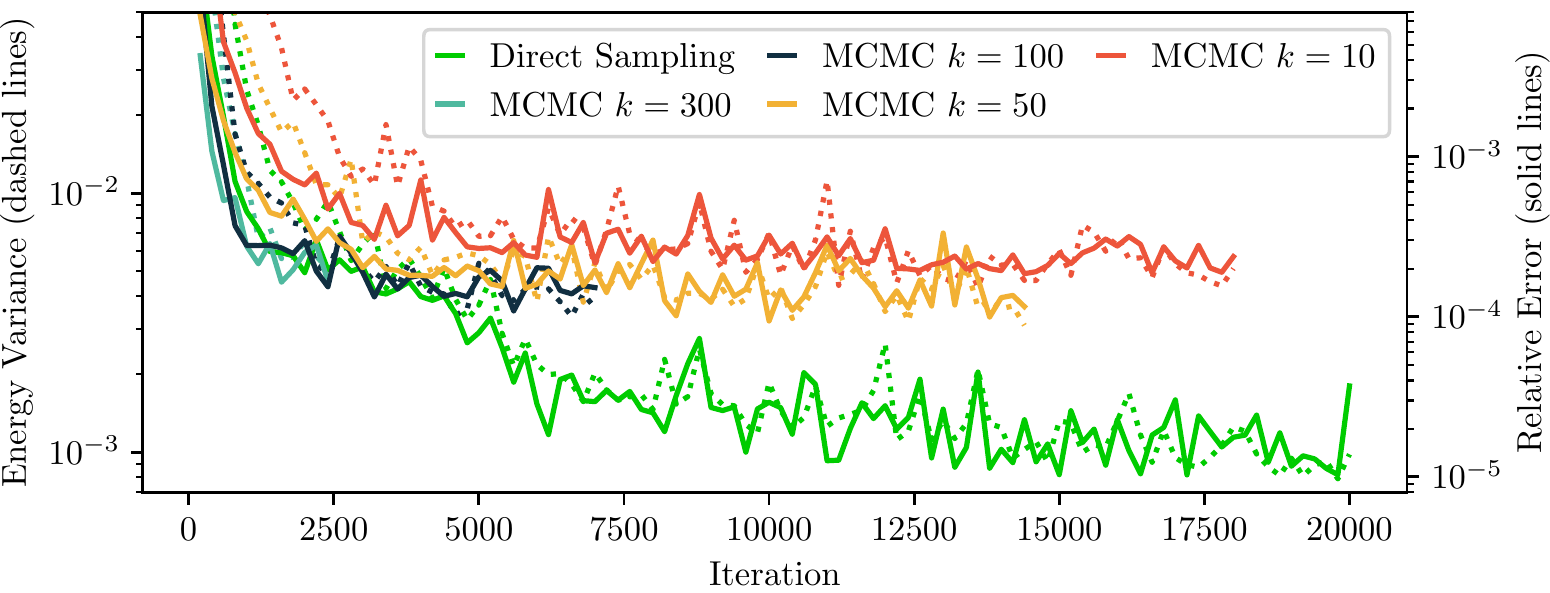}
    \caption{\label{fig:training_times}
    Comparing the effects of the
    sampling method, either MCMC or direct sampling, on the training
    procedure for the transverse-field Ising model with $\Gamma{=}3J$,
    close to the critical value, on a large ($21{\times}21$) lattice.
    When using MCMC, samples are taken every
    $k{\in}\left\{10, 50, 100, 300\right\}$ steps in the chain, where increasing
    $k$ reduces the correlation between samples at the expense of
    increased computational cost. The solid lines shows the relative
    error to the minimal energy found for this system in our
    experiments, and dashed lines shows the energy variance. Since MCMC
    takes a considerable time to complete just a single iteration, we
    have restricted the training to maximum of 100 hours.
    }
\end{figure}

In order to quantify the behavior of our model in a region of broken symmetry,
we consider the case of a transverse-field deep in the ferromagnetic region, namely
$\Gamma{=}2J$. The PCA visualization in Fig.~\ref{fig:broken_sym} shows that
for this value of $\Gamma$ the MCMC chains initialized at one of the oriented states composing the
ground state are stuck at that specific orientation and cannot come around to
sampling spin configurations that correspond to the opposite orientation.
In contrast, spin configurations sampled directly from the distribution by using
our proposed technique include equally probable configurations from both
orientations. 
The ergodicity breaking in local MCMC is also directly quantifiable 
by the expectation value of the total magnetization $m{\equiv}\left\langle\sum_i \sigma^z_i \right\rangle$,
for which we expect $m{=}0$ on any finite lattice. Indeed, the \iid sampling enabled
by our model correctly explores the two relevant ferromagnetic states (in agreement with
the visualization of Fig.~\ref{fig:broken_sym}) and reaches a value close to a total zero
magnetization, in stark contrast with MCMC estimation that effectively computes
$\left\langle\abs{\sigma_z}\right\rangle{\approx}0.78$ rather than $m$.
As expected, directly estimating $\left\langle\abs{\sigma_z}\right\rangle$ with our sampling method correctly
recovers it to a high precision, see Table~\ref{table:ground_state_values}.

The limitation of the MCMC procedure in providing independent samples is not only conceptually relevant, but
it can also have consequences on the quality of the resulting ground-state approximations.
In Fig.~\ref{fig:training_times}, we show the training procedure for the transverse-field 
$\Gamma{=}3J$, close to the critical value on a larger system ($21{\times}21$). 
The same NAQS architecture was trained once with the \iid sampling procedure and
once with MCMC chains of varying lengths. 
The optimization advantage obtained when relying on independent samples clearly
emerges from those figures~--~this procedure is much quicker and results in a
significantly more accurate ground state energy and
lower energy variance $\langle H^2 \rangle{-}\langle H \rangle ^2$.

\begin{table}
\begin{tabular}{cccc}
Lattice        &     PEPS     &     NAQS     &     QMC      \\ \hline
$10{\times}10$ & -0.628601(2) & -0.628627(1) & -0.628656(2) \\
$16{\times}16$ & -0.643391(3) & -0.643448(1) & -0.643531(2)
\end{tabular}
\caption{\label{table:heisenberg} Ground state energies for
    the antiferromagnetic Heisenberg model with open boundary
    conditions, as obtained by a state-of-the-art PEPS
    model~\citep{liu2017gradient}, our NAQS model, and the exact QMC
    estimation, as reported in \citet{liu2017gradient}.}
\end{table}

As a further benchmark, we also apply our method to a more complex system,
the two-dimensional antiferromagnetic Heisenberg model with open
boundary conditions, whose Hamiltonian is given by
$H{=}\sum_{\langle i,j\rangle}\sigma_x^i\sigma_x^j{+}\sigma_y^i\sigma_y^j{+}\sigma_z^i\sigma_z^j$.
We evaluate our approach by comparing the ground state energy obtained
for $10{\times}10$ and $16{\times}16$ systems with those obtained by QMC simulations, as
well as other variational methods. We find that NAQS meaningfully
improve upon the accuracy of the best known variational methods for this problem. 
Namely, for $10{\times}10$, a relative error of $8.7{\times}10^{-5}{\pm}0.6{\times}10^{-5}$ was
reported in \citet{liu2017gradient} using a PEPS model, whereas with
our approach we were able to obtain
$3.5{\times}10^{-5}{\pm}0.4{\times}10^{-5}$. See
table \ref{table:heisenberg} for exact results. While the PEPS results can be, in principle, further improved, 
increasing the accuracy comes with a very significant computational requirements~\cite{pepsplus} 
due to the unfavorable computational scaling \wrt the bond-dimension.
Moreover, though
not directly comparable, it is noteworthy that the relative error
of the ground state energy with \emph{periodic} boundary conditions
obtained by NQS with MCMC sampling is significantly less accurate than
ours (\citet{carleo2017solving,j1j2study} report relative error
greater than $2{\times}10^{-4}$).

\paragraph*{Discussion.--} 
In this work, we have shown a scheme to facilitate the practical employment of contemporary deep learning
architectures to the modeling of many-body 
quantum systems. This constitutes a striking improvement over currently used RBM
methods that are limited to only hundreds of parameters, and very shallow networks. 
A further practical advantage we gain is the ability to make use of the substantial body
of knowledge regarding optimization of these architectures that is accumulating
in the deep learning literature. 
We empirically demonstrate that by employing common deep learning optimization
methods such as SGD, our direct sampling approach
allows us to train very large convolutional networks ($20$~layers, $21{\times}21$~lattice, ${\sim}1M$~parameters).
Our presented experiments demonstrate that even for relatively simple systems MCMC
sampling can fail, while the \iid sampling enabled by our model succeeds.
Relying on the theoretically promising results regarding convolutional
networks' capabilities in representing highly entangled
systems~\cite{levine2019quantum}, namely, systems satisfying volume-law,
we view the enabling of their
optimization as an integral step in reaching currently unattainable
insight on a vast variety of quantum many body phenomena.

\begin{acknowledgments}
This work is supported by ISF Center grant 1790/12 and by the
European Research Council (TheoryDL project). Yoav Levine is
supported by the Adams Fellowship Program of the Israel Academy of
Sciences and Humanities. QMC simulations for the 2D Transverse-Field
Ising Model have been performed using the open-source ALPS
Library~\cite{bauer_alps_2011}.
\end{acknowledgments}

\ifdefined\ARXIV
   \appendix
   \section{Proof of Claim~\ref{claim:naqs_norm}}\label{app:proof}
The proof follows an induction argument. For $N=1$, it holds that
$\Psi(s_1) \equiv \Psi_1(s_1)$, and so $\Psi$ is normalized because $\Psi_1$
is normalized with respect to $s_1$. Assume the claim holds for $N=k$, then
for $N=k+1$ we first define $\tilde{\Psi}(s_1,\ldots,s_k) \equiv \prod_{i=1}^k \Psi_i(s_i | s_{i-1},\ldots,s_1)$,
and so
\begin{align*}
    &\sum_{s_1,\ldots,s_{k+1}} \abs{\Psi(s_1,\ldots,s_{k+1})}^2 \\
    &= \sum_{\mathclap{s_1,\ldots,s_{k+1}}} \quad\prod_{i=1}^{k+1} \abs{\Psi_i(s_i | s_{i-1},\ldots,s_1)}^2 \\
    &= \sum_{\mathclap{s_1,\ldots,s_k}} \,\left(\prod_{i=1}^k \abs{\Psi_i(s_i | s_{i-1},\ldots,s_1)}^2\right)\!\! \overbrace{\sum_{s_{k+1}} \abs{\Psi_i(s_{k+1} | s_k,\ldots,s_1)}^2}^{*=1} \\
    &= \sum_{\mathclap{s_1,\ldots,s_k}} \tilde{\Psi}(s_1,\ldots,s_k) \overset{**}{=} 1,
\end{align*}
where $(*)$ is because $\Psi_{k+1}$ is a normalized conditional wave function, and $(**)$ because of the induction assumption.
\hfill $\square$ 


\section{Technical Details}\label{app:tech}
In this section we cover the essential technical details of our models and how they are optimized.

\subsection{Architecture}

Our chosen architecture for our implementation of Neural Autoregressive Quantum State is loosely inspired by that of
PixelCNN~\citep{van2016conditional}, which uses a row-wise ordering of the particles for the conditional wave-functions.
All parameters and operations in the network are real, where the complex log-amplitudes of the conditional wave functions
are represented as two real numbers, as discussed in the body. More specifically, it is composed of two interacting branches:
(i)~a ``vertical'' branch for representing conditional dependencies between a given particle
and all particles above it in the 2D lattice it resides on, and (ii)~a ``horizontal'' branch for representing conditional
dependencies between a given particle and all particles to its left.
Each branch comprises a sequence of convolutional layers. The ``vertical'' containing convolutional layers with
$3 \times 3$ filters and 32 channels, where we add two rows of zero-padding to the top lattice before applying the convolution, to
ensure a particle is not dependent on rows below it. For the ``horizontal'' branch, we use convolutional layers with
$3 \times 3$ filters and 32 channels, where we add two rows and two columns of zero-padding to the top and left of the
lattice before applying the convolution, as well as setting the parameters at the $(3,3)$ indices to be zero, all to ensure
that a particle only depends on particles not ``ahead'' of itself according to the row-wise ordering we enforce. To combined
them, the two branches are connected using the following scheme: after every convolutional layer in the ``vertical'' branch,
but before the convolutional layer of the ``horizontal'' branch, we take the intermediate result of the ``vertical'' branch,
shift every entry ``down`` along the vertical axis of the lattice, and concatenate it along the channels axis of the horizontal branch.
The final convolutional layer of the ``horizontal'' branch serves as the output of the network, and hence we only use 2 output channels
in that final layer, the two coordinates serving as the real and imaginary parts of the log-amplitude of the conditional wave-functions.
The complete network can be depicted as a sequence of blocks, each as illustrated in fig.~\ref{fig:network_block}.
We typically use between 10 and 40 such blocks, depending on the specific experiment. This separation of ``vertical'' and
``horizontal'' branches is a technique to overcome what is known as the ``blind spot'' problem of the original PixelCNN
architecture~(see \citet{van2016conditional} for more details).

\begin{figure}
    \centering
    \includegraphics[width=1\linewidth]{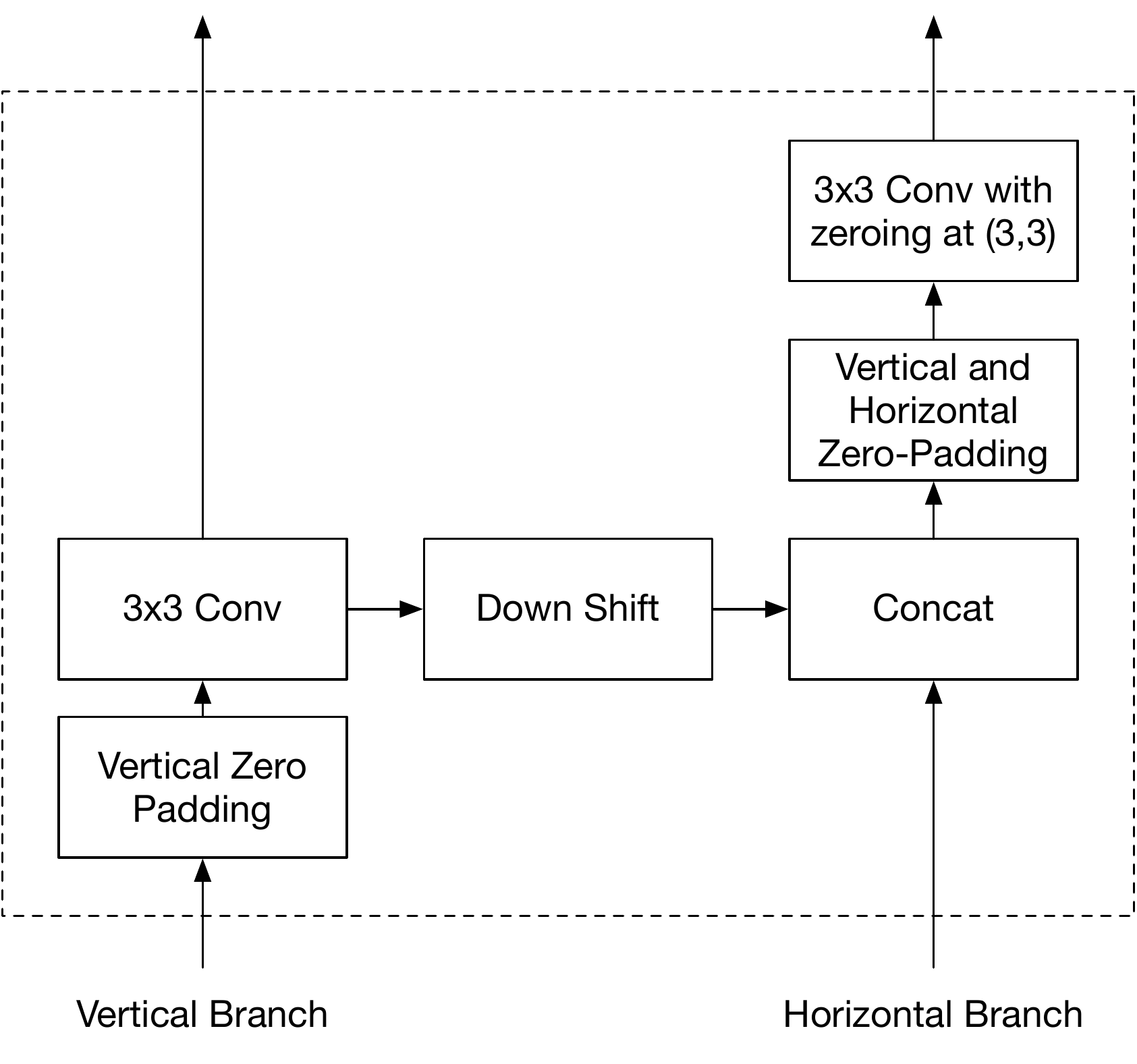}
    \caption{\label{fig:network_block}An illustration of a single block that our architecture is composed of.
    }
\end{figure}

\subsection{Handling Symmetries}

While our general NAQS architecture can already represent wave-functions quite well,
we have found that leveraging the inherent symmetries of a given problem, e.g.,
invariance to rotations and flips, can dramatically improve the accuracy of our model.
Specifically, we use a self-ensemble scheme to \emph{symmetrize} our model, where for a
model $f(\s)$ and a given input spin-configuration, we transform it according to its
symmetries, denoted by the set $\mathcal{T}$, run each of them through our model, and
aggregate the resulting log-amplitude outputs using the following equations for our
symmetrize model $\mathrm{Sym}(f)$:
\begin{align}
    \mathrm{Re}(\mathrm{Sym}(f)(\s)) &{=} \frac{1}{2}{\ln}\left(\sum_{T\in \mathcal{T}} \frac{1}{\abs{\mathcal{T}}} e^{2\cdot\mathrm{Re}(f(T\s))}\right), \label{eq:ensemble_real} \\
    \mathrm{Im}(\mathrm{Sym}(f)(\s)) &{=} {\mathrm{Im}}\!\left(\!{\ln}\!\left(\sum_{T\in \mathcal{T}} \e^{i\cdot\mathrm{Im}(f(T\s))}\right)\!\right). \label{eq:ensemble_im}
\end{align}
It is important to emphasize that while there are many ways to symmetrize a model, we
cannot use any aggregation operation~--~it must also preserve its probabilistic meaning
for we to be able to sample from it efficiently. We propose to incorporate the possible
symmetries into our generative model, assuming we first sample a transformation $T$ from
$\mathcal{T}$ with equal probability $\nicefrac{1}{\mathcal{T}}$, and then draw a sample
from our model as described in the main text, followed by transforming it with $T$. This
translates to a mixture model over the squared magnitudes of the network's predicted
amplitude, and eq.~\ref{eq:ensemble_real} realizes it in log-space, where the real part
of the output represent the log-magnitude. For the imaginary part, we have less
restrictions and most symmetric operators would work, but we found that the
\emph{mean of circular quantities} of the phases, as expressed in
eq.~\ref{eq:ensemble_im}, worked best in our experiments.

\subsection{Optimization}

In our experiments we employ the following general optimization strategy. We begin using the Adam~\citep{kingma2014adam}
SGD variant, using a small batch of 100 samples for estimating the gradient and using a learning
rate in the order of $10^{-3}$. After about 10K gradient update steps, we increase the batch size to 1000,
using the same learning rate and optimizer as the first stage, for an additional 10K update steps. In the final stage
of the optimization, we increase the batch size again, and also switch to standard SGD with a momentum term, for
an additional 5K update steps. For each experiment, we test multiple variations around the above default values of
batch size, learning rate, and number of update steps in each stage, and report the results for the best performing models.

\fi

\bibliography{refs}

\end{document}